\documentstyle[prl,aps,multicol]{revtex}

\begin{document}
\draft

\title{ Quantum melting of magnetic order due to orbital fluctuations }

\author { Louis Felix Feiner }
\address{ Philips Research Laboratories, Prof. Holstlaan 4, 
          NL-5656 AA Eindhoven, The Netherlands }
\author { Andrzej M. Ole\'{s}\cite{AMO} }
\address{ Max-Planck-Institut f\"ur Festk\"orperforschung,
       Heisenbergstrasse 1, D-70569 Stuttgart, Federal Republic of Germany}
\author { Jan Zaanen }
\address{ Lorentz Institute for Theoretical Physics, Leiden University,
          P.O.B. 9506, NL-2300 RA Leiden, The Netherlands }
\date{November 27, 1996}
\maketitle

\begin{abstract}
We have studied the phase diagram and excitations of the spin-orbital 
model derived for a three dimensional perovskite lattice, as in KCuF$_3$. 
The results demonstrate that the orbital degeneracy drastically increases 
quantum fluctuations and suppresses the classical long-range order near 
the multicritical point in the mean-field phase diagram. 
This generates a qualitatively new spin liquid, providing the first example 
of a valence bond ground state in three dimensions.
\end{abstract}


\begin{multicols}{2} 

It is common wisdom that macroscopic ensembles of interacting particles 
tend to behave classically. This is not always true, however, and the study 
of collective quantum systems starts to become a prominent 
theme in condensed matter physics. Central to this pursuit are 
low-dimensional quantum spin systems (spin chains \cite{Whi94} and ladders
\cite{Dag96}), and it proves difficult to achieve quantum melting of 
magnetic long-range order (LRO) in empirically relevant systems in higher 
dimensions. Here we want to suggest a class of systems in which 
quantum-melting occurs even in three dimensions: small spin, orbital 
degenerate magnetic insulators, the so-called Kugel-Khomskii (KK) systems 
\cite{Kug82}. There might exist already a physical realization of such 
a three-dimensional (3D) "quantum spin-orbital liquid": LiNiO$_2$. 

Global $SU(2)$ by itself is not symmetric enough to defeat classical order
in $D>1$ and the pursuit has been open for some time to engineer more
fluctuations into these systems. Three (related) strategies to realize
quantum melting have proven to be successful: 
(i) Adding zero-dimensional fluctuations as in the bilayer Heisenberg model 
which leads to an incompressible spin liquid \cite{Mil94,Chu95}.
(ii) Frustrating the system so that the classical sector gets highly 
degenerate, as in the case of the $S=1/2$ square lattice with longer ranged 
antiferromagnetic (AF) interactions ($J_1$-$J_2$-$J_3$ models 
\cite{Pre88,Chu91}). These systems involve fine-tuning of parameters and 
are therefore hard to realize by chemistry. 
(iii) Finally, reducing the number of magnetic bonds, as in the 1/5 depleted 
square lattice, where the resulting plaquette resonating valence bond (PRVB)
state explains the spin gap observed in CaV$_4$O$_9$ \cite{Ued96}. 
In this Letter we show that orbital degeneracy operates through the same
basic mechanisms to produce quantum melting in the KK systems. The 
novelty is that these systems tend to "self-tune" to (critical) points of 
high classical degeneneracy. There are interactions which may lift the 
classical degeneracy, but they are usually weak. 

An interaction of this kind is the electron-phonon coupling -- the 
degeneracy is lifted by a change in crystal structure, the conventional 
collective Jahn-Teller (JT) instability. However, as was pointed out in the 
seminal work by Kugel and Khomskii \cite{Kug82}, in orbital 
degenerate Mott-Hubbard insulators one has to consider in first instance 
the purely electronic problem. Because of the large local Coulomb 
interactions (Hubbard $U$), a low energy Hilbert space splits off, spanned  
by {\it spin and orbital} configuration space, with superexchange-like 
couplings between both spin and orbital local degrees of freedom. The 
orbital sector carries a discrete symmetry and the net outcome is that the 
clock-like orbital degrees of freedom get coupled into the $SU(2)$ spin 
problem. Such a system might undergo symmetry breaking in states with 
simultaneous spin- and orbital order. The lattice has to react to the 
symmetry lowering in the orbital sector, but it was recently convincingly
shown, at least in the archetypical compound KCuF$_3$, that the structural 
distortion is a side effect \cite{Lie95}.

The fundamental question arises if these forms of classical order are 
always stable against quantum fluctuations. Although the subject is much
more general (singlet-triplet models in rare earth compounds \cite{Hsi72},
V$_2$O$_3$ \cite{Cas78}, LaMnO$_3$ \cite{Miz96}, heavy fermions 
\cite{Cox87}), we focus here on the simplest situation encountered in 
KCuF$_3$ and related systems \cite{Kug82}. These are JT-distorted cubic, 
3D analogues of the cuprate superconductors \cite{Web88}. The magnetic ion 
is in a $3d^9$ state, characterized in the absence of JT-distortion by two 
degenerate $e_g$ ($x^2-y^2 \sim |x\rangle$, $3z^2-1 \sim |z\rangle$) 
orbitals, next to the $S=1/2$ spin degeneracy. Kugel and Khomskii derived 
the effective Hamiltonian \cite{Kug82} with AF superexchange $J=t^2/U$ 
(where $t$ is the hopping between $|z\rangle$ orbitals along the $c$-axis), 
\begin{eqnarray} 
\label{kk1}
H_1 = &J& \sum_{\langle ij\rangle,\alpha} \left[ 4(\vec{S}_i\cdot\vec{S}_j ) 
  (\sigma^{\alpha}_i - \frac{1}{2}) (\sigma^{\alpha}_j - \frac{1}{2})\right.
                                   \nonumber \\ 
& & \hskip 1.0cm + \left. (\sigma^{\alpha}_i+\frac{1}{2})(\sigma^{\alpha}_j 
+ \frac{1}{2}) - 1 \right] ,
\end{eqnarray} 
neglecting the Hund's rule splittings of the intermediate $d^8$ states. 
Including those up to order $\eta=J_H/U$ ($J_H$ is the singlet-triplet 
splitting) yields in addition,
\begin{eqnarray} 
\label{kk2}
H_2 = & J\eta & \sum_{\langle ij\rangle,\alpha}
  \left[ (\vec{S}_i\cdot\vec{S}_j) 
  (\sigma^{\alpha}_i + \sigma^{\alpha}_j - 1 )  \right.   \nonumber \\
&+& \left. \frac{1}{2}(\sigma^{\alpha}_i-\frac{1}{2}) 
                      (\sigma^{\alpha}_j-\frac{1}{2}) 
 + \frac{3}{2} (\sigma^{\alpha}_i \sigma^{\alpha}_j - \frac{1}{4})\right] . 
\end{eqnarray}
In Eqs. (\ref{kk1}) and (\ref{kk2}), $\vec{S}_i$ refers 
to the spin at site $i$, while $\alpha$ refers to the cubic ($a,b,c$) axes, 
and the orbital degrees of freedom are represented by
\begin{equation} 
\label{orbop}
\sigma^{a(b)}_i = \frac{1}{4}( -\tau^z_i \pm\sqrt{3}\tau^x_i ), \hskip .7cm
\sigma^c_i = \frac{1}{2} \tau^z_i.
\end{equation}
The $\tau$'s are Pauli matrices acting on the orbital pseudo-spins
$|x\rangle ={\scriptsize\left( \begin{array}{c} 1\\ 0\end{array}\right)},\; 
 |z\rangle ={\scriptsize\left( \begin{array}{c} 0\\ 1\end{array}\right)}$. 
Hence, we find a Heisenberg Hamiltonian for the spins, coupled into an 
orbital problem which is clock-model-like (there are three directional 
orbitals: $3x^2-1$, $3y^2-1$, and $3z^2-1$, but they are not independent)
\cite{notehitc}. As we will see in a moment, the Hund's rule coupling term 
(\ref{kk2}) acts to lift the degeneracy. Next to this, we introduce another 
control parameter,
\begin{equation} 
\label{kk3}
H_3 = - E_z \sum_i \sigma^c_i,
\end{equation}
a "magnetic field" for the orbital pseudo-spins, loosely associated with 
an uniaxial pressure along the $c$-axis.

The classical phase-diagram of the spin-orbital model $H=H_1+H_2+H_3$, 
shown in Fig. 1, demonstrates the competition between the spin and orbital 
interactions similar to that found before in two dimensions \cite{crete}. 
It consists of five phases with staggered (two-sublattice) magnetic
long-range order (LRO): 
(i) At large positive $E_z$, the orbital system is uniformly polarized 
along $x^2-y^2$. As no superexchange is possible in the $c$-direction, the 
$(a,b)$ planes decouple magnetically, and we recognize the two-dimensional 
(2D) antiferromagnet, called further AFxx, well known from the cuprate 
superconductors. 
(ii) At large negative $E_z$, the orbitals polarize along $3z^2-1$, and 
the spin system is an anisotropic 3D antiferromagnet, called AFzz. These 
two AF phases (AFxx and AFzz) are stabilized by the anisotropic superexchange 
which amounts to $4J$ between $z$ orbitals along the $c$-axis, and to $9J/4$ 
($J/4$) between the $x$ ($z$) orbitals in the $(a,b)$-planes, respectively. 
In contrast, the Hund's rule $J_H$ stabilizes mixed-orbital (MO) phases in 
which both AF and ferromagnetic (FM) couplings occur: 
(iii) At large $J_H/U$ and $E_z<0$, a MO{\scriptsize FFA} phase 
is found, characterized at each site by orbitals, 
$|i\sigma\rangle=\cos\theta_i |x\sigma\rangle+\sin\theta_i |z\sigma\rangle$,
with the sign of $\theta_i$ alternating between the two sublattices in the 
$(a,b)$ planes. At small $|E_z|$ the orbitals stagger like $x^2-z^2$,
$y^2-z^2$, $x^2-z^2, \cdots$, and point towards each other along the
$c$-axis which results there in strong ($\sim J$) AF interactions. In 
contrast, weak FM interactions ($\sim J_H$) occur within the $(a,b)$ planes. 
(iv) A similar MO state, called MO{\scriptsize AFF}, is found
at large $J_H/U$ and $E_z>0$, 
with the directional orbitals staggered in FM $(b,c)$ planes, and AF order
along the $a$ direction \cite{moaff}. Both states are degenerate 
at the line $E_z=0$, where $\cos 2\theta_i=-(1-\frac{\eta}{2})/(2+3\eta)$, 
and one is close to the KK phase \cite{Kug82} in KCuF$_3$, where, 
within the limitations of chemistry, the magnetic order appears to be 
equivalent to an ideal one-dimensional (1D) $S=1/2$ spin system. 
(v) Finally, the MO{\scriptsize AAF} phase with a small
$|z\rangle$-amplitude,
AF order in the $(a,b)$ planes and FM order along the $c$-axis,
is stable between the MO{\scriptsize AFF} and AFxx phases. 
Thus, we find the same ingredient as in the frustrated Heisenberg 
antiferromagnet (HAF): critical lines where different classical spin 
structures become degenerate. The present case appears to be more extreme 
as even the effective dimensionality of the spin system changes because of 
the coupling to the orbital sector. The frustration manifests itself at the 
classical degeneracy point $M\equiv (E_z,\eta)=(0,0)$, where the orbitals 
may be rotated freely when the spins are AF, and the spins may be rotated 
freely within the FM planes of the MO phases. The same energy of $-3J$ per 
site is obtained either in a 3D antiferromagnet with completely frustrated 
orbitals [consider $\langle\vec{S}_i\vec{S}_{i+\delta}\rangle=-1/4$ in Eq. 
(\ref{kk1})], or in a disordered spin system due to the orbital sector.

It is instructive to consider the stability of the classical phase diagram
to Gaussian quantum fluctuations \cite{crete}. The collective modes can be 
calculated using, e.g., a random phase approximation (RPA) within the Green 
function technique \cite{Hal72}. Next to the Goldstone modes of the spin 
system, one finds optical modes corresponding to orbital excitations which 
occur both in the presence ('transverse') and in the absence 
('longitudinal') of a simultaneous spin-flip. The new feature is that the 
{\em spin and transverse orbital excitations are coupled}, so that 
fluctuations in the orbital sector also affect the spin sector. The 
approach of the critical lines is signalled by the softening of both 
longitudinal and transverse orbital modes. The (mixed) transverse modes 
give the dominating contribution to the renormalization of energy and 
magnetic order parameter. In the AFxx (AFzz) phase the lowest transverse 
mode softens along $\vec{k}=(\pi,0,k_z)$ [$\vec{k}=(k_x,0,0)$], and 
equivalent lines in the Brillouin zone (BZ), regardless how one approaches 
the critical lines. Thus, these modes become dispersionless along 
particular (soft-mode) lines in the BZ, where we find {\em finite} masses 
in the perpendicular directions,
\begin{eqnarray} \label{mass0}
\omega_{\rm AFxx}(\vec{k}) \rightarrow & \Delta_x &
      + B_x \left( k_x^4 + 14k_x^2k_y^2 + k_y^4 \right)^{1/2}, \nonumber \\
\omega_{\rm AFzz}(\vec{k}) \rightarrow & \Delta_z & 
      + B_z \left( k_y^2 + 4k_z^2 \right), 
\end{eqnarray}
with $\Delta_i=0$ and $B_i\neq 0$ at the $M$ point, and the quantum 
fluctuations diverge logarithmically, $\langle\delta S\rangle\sim
\int d^3k/\omega(\vec{k})\sim\int d^2k/(\Delta_i+B_ik^2)\sim\ln\Delta_i$, 
if $\Delta_i\rightarrow 0$ at the transition. An analytic expansion could
not be performed in the MO phases, but the numerical results reported 
below suggest a qualitatively similar behavior.

We have verified that the above behavior of the soft mode results in large 
quantum corrections $\langle\delta S^z\rangle$ to the order parameter in 
all magnetic phases close to the critical lines. As an example we show
$\langle S^z\rangle$ in AFxx and AFzz phases, being significantly lower 
than in a 2D HAF in a broad parameter regime (Fig. 2). Similarly as in 
two dimensions \cite{crete}, the LRO is overwhelmed by quantum fluctuations 
at particular lines, where $|\langle \delta S^z\rangle|=\langle S^z\rangle$ 
(Fig. 1). Unlike $\langle \delta S^z\rangle$, the RPA energies of the 
ordered phases show no divergence, with quite similar energy gains in AF 
and MO phases of the order of $0.6J$ \cite{rpamo}. 
We therefore believe that RPA is here as accurate as in the pure-spin HAF 
and conclude that the {\it classical order is destroyed by quantum 
fluctuations} in the small $|E_z|$ and small $\eta$ region between dashed 
and dotted lines in Fig. 1. 

Although the above theory is known to perform quite well in the simplest 
systems \cite{Cha88}, it might be misleading in more complicated 
situations. For instance, the "finite mass mode softening" occurs also in 
the $J_1$-$J_2$-$J_3$ model where it is shown to be inconsequential in the 
large $S$ limit because of an "order out of disorder" phenomenon 
\cite{Chu91}. In contrast, low order spin-wave theory is blind for the 
quantum transition occurring in the bilayer model \cite{Chu95}. In all 
these cases, including ours, the quantum melting is promoted by the drastic 
enhancement of {\em local} fluctuations. It is then instructive to consider 
valence bond (VB) states \cite{Aue94} with the individual spins paired into 
singlets and the orbitals optimized variationally. As the energy of 
a singlet is lowest when the orbitals point along the bond, the optimal 
states with all singlets lined up in parallel (see Fig. 3) are: 
(i) for $E_z>0$, singlets along the $a$-axis with orbitals close to 
$3x^2-1$ (VBa), and (ii) for $E_z<0$, singlets along the $c$-axis with 
orbitals $\sim |z\rangle$ (VBc). Both optimize spin and orbital 
energy on every second bond, and have lower energy than the classical 
states close to the classical degeneracy \cite{notevb}. 

Further, we included the leading quantum fluctuations in the VB states. 
The energy of the resonating VBc (RVBc) state was obtained using the Bethe 
ansatz result for the 1D HAF. We attempted to improve the VBa state by 
constructing the PRVB states, $|\Psi_{\Box}\rangle\sim 
(|\Psi_a\rangle +e^{i\phi}|\Psi_b\rangle)$, from the singlet pairs along 
$a$ and $b$, $|\Psi_a\rangle$ and $|\Psi_b\rangle$. Surprisingly, more 
energy is gained instead in the plaquette VB (PVB) states in which the 
wave functions $|\Psi_a\rangle$ and $|\Psi_b\rangle$ alternate and form a 
superlattice. The exceptional stability of these (nonresonating) PVB states 
is due to a unique mechanism involving the orbital sector. Unlike 
in the HAF, the bonds not occupied by the singlets {\em contribute orbital 
energy} and this is optimized when singlets in orthogonal directions are 
connected \cite{notevb}. If $\eta<0.30$, this PVB alternating (PVBA) state 
(Fig. 3) is stable at $E_z>0$, while a similar PVB interlayered state with 
alternating layers of $(a,b)$-plane/$c$-axis bonds (PVBIc), and the RVBc 
state are stable at $E_z<0$. Finally, a state interlayered along $a$, but in 
the same pattern as in PVBIc, occurs in between the PVBA and PVBIc states.
Thus, a spin liquid is stabilized by the {\it orbital degeneracy} over the
MO phases with RPA fluctuations in a broad regime (Fig. 4). This resembles 
the situation in a 2D 1/5-depleted lattice \cite{Ued96}, but the present 
instability is much stronger and happens in {\em three} dimensions.

Summarizing, we find strong theoretical arguments supporting the 
conjecture that quantum-melting might occur in orbital degenerate
Mott-Hubbard insulators. Why does it not occur always (e.g., in KCuF$_3$)? 
Next to the Hund's rule coupling, the electron-phonon coupling $\lambda$ 
is dangerous. The lattice will react to the orbital fluctuations, dressing 
them up in analogy with polaron physics, and thereby reducing the coupling 
constant. In order to quantum melt KCuF$_3$-like states, one should 
therefore look for ways to reduce both the effective $J_H$ and $\lambda$. 
We believe that this situation is encountered in LiNiO$_2$: although the 
spin-spin interactions in the (111) planes should be very weakly FM 
according to the Goodenough-Kanamori rules, magnetic LRO is absent 
\cite{Hir85} and the system might represent the spin-orbital liquid. More 
strikingly, LiNiO$_2$ is cubic and should undergo a collective JT 
transition, which absence is actually an old chemistry mystery!
Upon electron-hole transformation, $d^7$ low-spin Ni$^{3+}$ maps 
on $d^9$ Cu$^{2+}$ in KCuF$_3$, but with a difference in chemistry. 
While the $e_g$ hole in KCuF$_3$ is nearly entirely localized on the Cu, 
the $e_g$ electron in LiNiO$_2$ is rather strongly delocalized over 
the Ni and surrounding O ions which reduces both $J_H$ and 
$\lambda$, and explains the absence of classical ordering. A more 
precise experimental characterization of LiNiO$_2$ is needed.

We thank D. I. Khomskii, M. Takano, and P. Horsch for stimulating
discussions, and acknowledge the support by the Committee of Scientific 
Research (KBN) of Poland, Project No. 2 P03B 144 08 (AMO), by the 
Dutch Academy of Sciences (KNAW) (JZ), and by ISI Foundation 
and EU PECO Network ERBCIPDCT940027.

\narrowtext

\begin{figure}
\caption
{Mean-field phase diagram of the spin-orbital model (1-4) in the
$(E_z,J_H)$ plane. Full lines indicate transitions between the 
classical states, while LRO is destroyed above the dashed (below the dotted) 
lines for the AF (MO) phases.}
\label{phdmfa}
\end{figure}

\begin{figure}
\caption
{Renormalization of AF LRO $\langle S^z_i\rangle$ in AFzz (left) and AFxx
(right) phases as functions of $E_z/J$.}
\label{S_z}
\end{figure}

\begin{figure}
\caption
{Schematic representation of spin singlets (double lines) in the disordered 
states: VBa, VBc, and PVBA. In the PVBA state the plaquettes with singlets 
$\parallel a$ and $\parallel b$ alternate both in the $(a,b)$ planes and 
in the $c$ direction, while the PVBIc state is obtained by interleaving
successive PVBA $(a,b)$ planes with one VBc double layer.}
\label{energy}
\end{figure}

\begin{figure}
\caption
{The same as in  Fig. 1, but including quantum fluctuations. 
The spin liquid (RVBc, PVBIc, PVBIa, and PVBA) 
takes over in the shaded crossover regime.}
\label{phdqf}
\end{figure}

\end{multicols} 


\begin{references}
\bibitem{moaff}
              The MO{\scriptsize AFF} phase is degenerate with a
              MO{\scriptsize FAF} phase which has
              the orbitals staggering within the $(a,c)$ planes.
\bibitem{rpamo}
              The energy in the MO phases for $E_z>0$ was estimated using 
              the averaged crystal field over the two sublattices.
\bibitem{notevb}
              Bonds not occupied by the spin singlets (dashed lines in 
              Fig. 3) contribute with finite orbital energies which depend 
              on the topology of singlet bonds. 
              They were included in the disordered states.
\end{references}
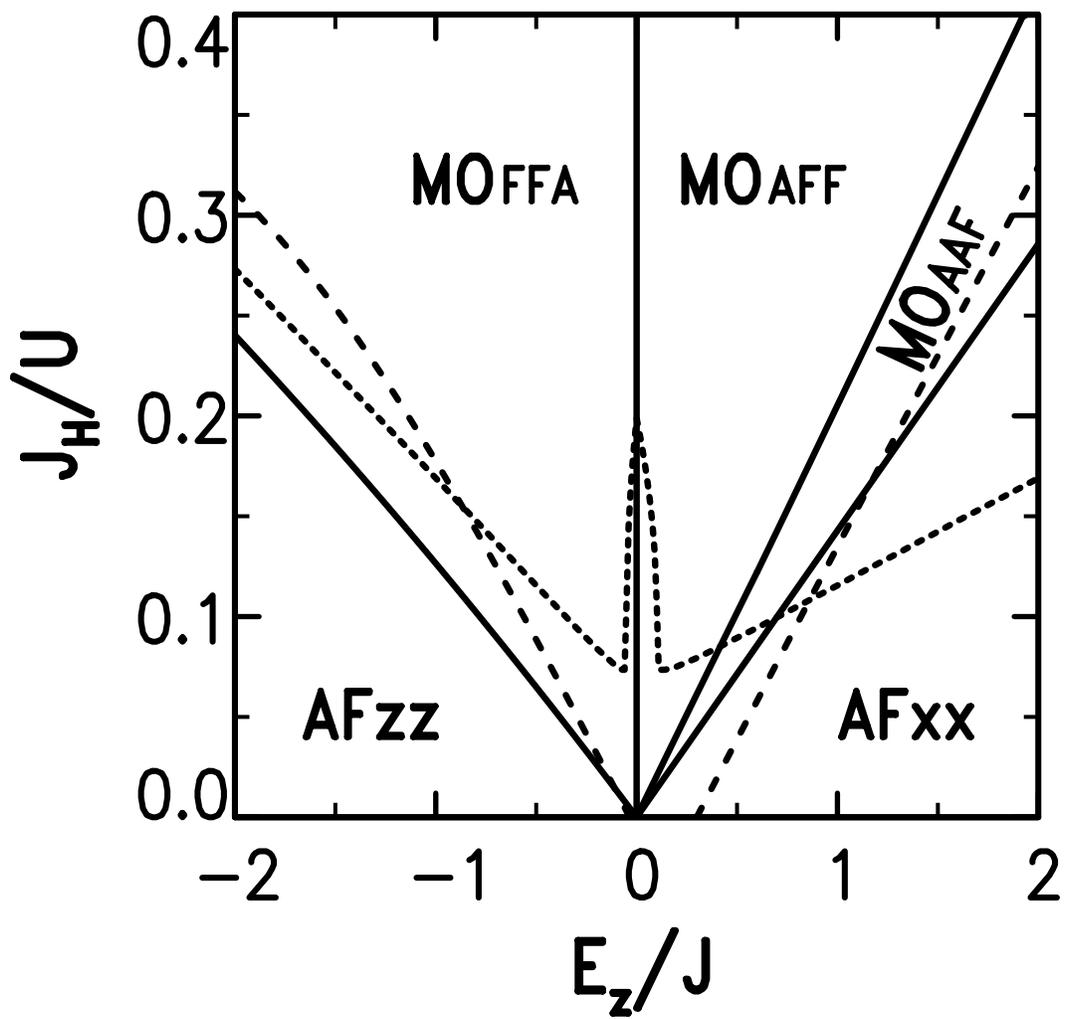
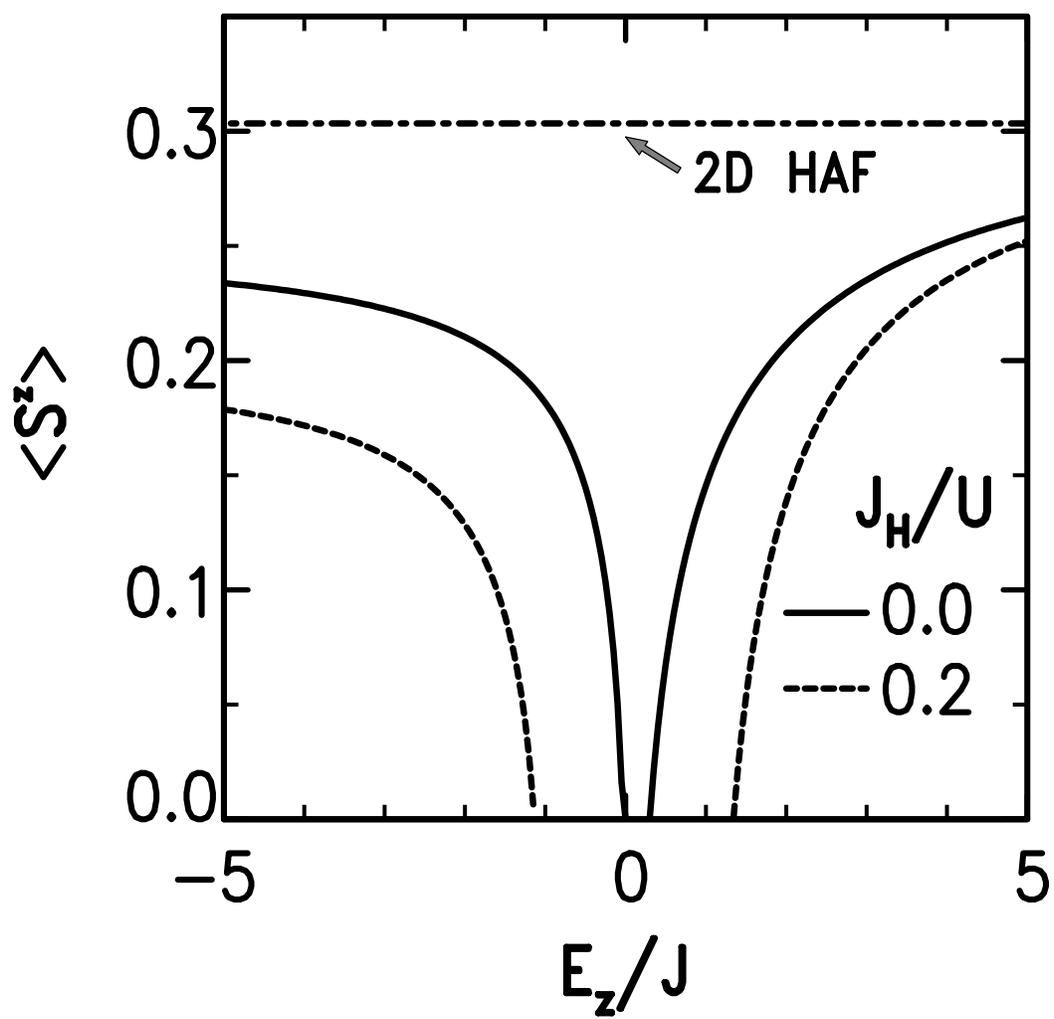
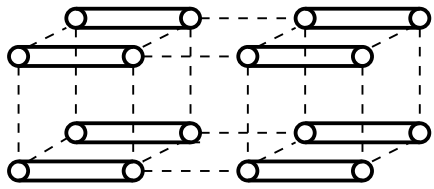
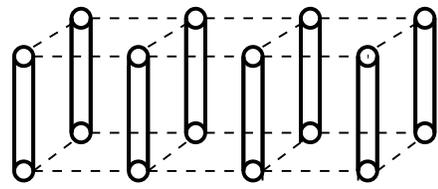
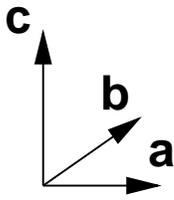
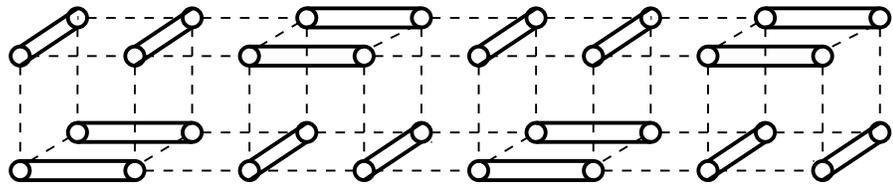
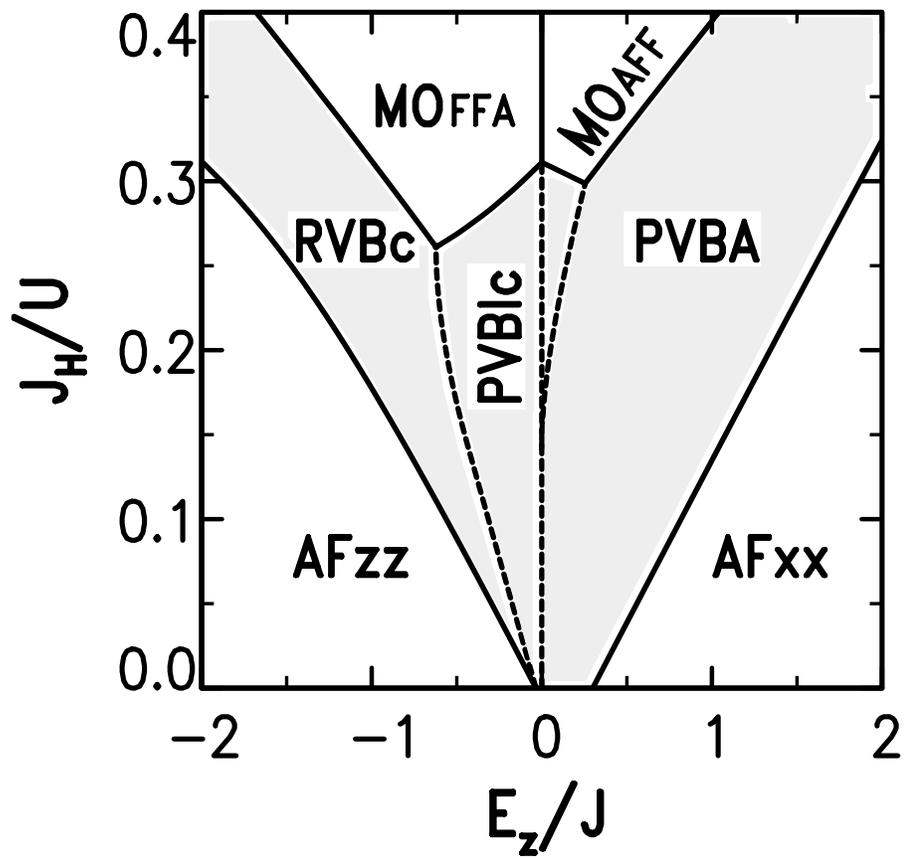
\end{document}